\documentclass{nature}

\usepackage{comment}

\usepackage{amsmath}
\usepackage[dvipsnames]{xcolor}
\usepackage{soul}
\usepackage{graphicx}

\usepackage[switch, modulo]{lineno}
\modulolinenumbers[1]

\newcommand{\Teff}{\mbox{$T_{\mathrm{eff}}$}}

\newcommand{\sdss}[3]{SDSS\,J#1$#2$#3}

\newcommand{\Msun}{\mbox{$\mathrm{M}_{\odot}$}}

\newcommand{\kms}{\mbox{$\mathrm{km\,s^{-1}}$}}
\newcommand{\masy}{mas\,yr$^{-1}$}
\newcommand{\pcm}{\mbox{cm$^{-1}$}}
\newcommand{\gcc}{\mbox{g\,cm$^{-3}$}}

\newcommand{\Ion}[2]{#1{\,\sc#2}}
\newcommand{\logXY}[2]{\mbox{$\log(\mathrm{#1/#2})$}}

\newcommand{\Lia}{LHS\,2534}
\newcommand{\Lib}{WD\,J2317$+$1830}
\newcommand{\Lic}{WD\,J1824$+$1213}
\newcommand{\Lid}{\sdss{1330}{+}{6435}}
\newcommand{\LibLong}{WD\,J231726.74$+$183052.75}
\newcommand{\LicLong}{WD\,J182458.45$+$121316.82}
\newcommand{\LidLong}{\sdss{133001.17}{+}{643523.69}}

\title{Alkali metals in white dwarf atmospheres as tracers of\linebreak ancient planetary crusts}

\author{Mark~A.~Hollands$^{1}\star$,
Pier-Emmanuel~Tremblay$^{1}$,
Boris~T.~G\"ansicke$^{1,2}$,
Detlev~Koester$^{3}$,\\
Nicola~Pietro~Gentile-Fusillo$^{4}$}

\begin{document}

\maketitle

\begin{affiliations}
 \item Department of Physics, The University of Warwick, Coventry, CV4 7AL, UK
 \item Centre for Exoplanets and Habitability, University of Warwick, Coventry, CV4 7AL, UK
 \item Institut f\"ur Theoretische Physik und Astrophysik, University of Kiel,
       24098 Kiel, Germany
 \item European Southern Observatory, Karl-Schwarzschild-Str 2, D-85748 Garching, Germany
\end{affiliations}

\begin{abstract}
White dwarfs that accrete the debris of tidally disrupted asteroids\cite{jura03-1} provide the opportunity to measure the bulk composition of the building blocks, or fragments, of exoplanets\cite{zuckermanetal07-1}. This technique has established a diversity in compositions comparable to what is observed in the solar system\cite{gaensickeetal12-1}, suggesting that the formation of rocky planets is a generic process\cite{doyleetal19-1}. Whereas the relative abundances of lithophile and siderophile elements within the planetary debris can be used to investigate whether exoplanets undergo differentiation\cite{zuckermanetal11-1}, the composition studies carried out so far lack unambiguous tracers of planetary crusts\cite{bonsoretal20-1}. Here we report the detection of lithium in the atmospheres of four cool ($<5,000$\,K)  and old (cooling ages $5$--$10$\,Gyr) metal-polluted white dwarfs, where one also displays photospheric potassium. The relative abundances of these two elements with respect to sodium and calcium strongly suggest that all four white dwarfs have accreted fragments of planetary crusts. We detect an infrared excess in one of the systems, indicating that accretion from a circumstellar debris disk is on-going. The main-sequence progenitor mass of this star was $4.8\pm0.2$\,\Msun, demonstrating that rocky, differentiated planets may form around short-lived B-type stars. 
\end{abstract}

\begin{figure}
  \centering
  \includegraphics[angle=0,width=\textwidth]{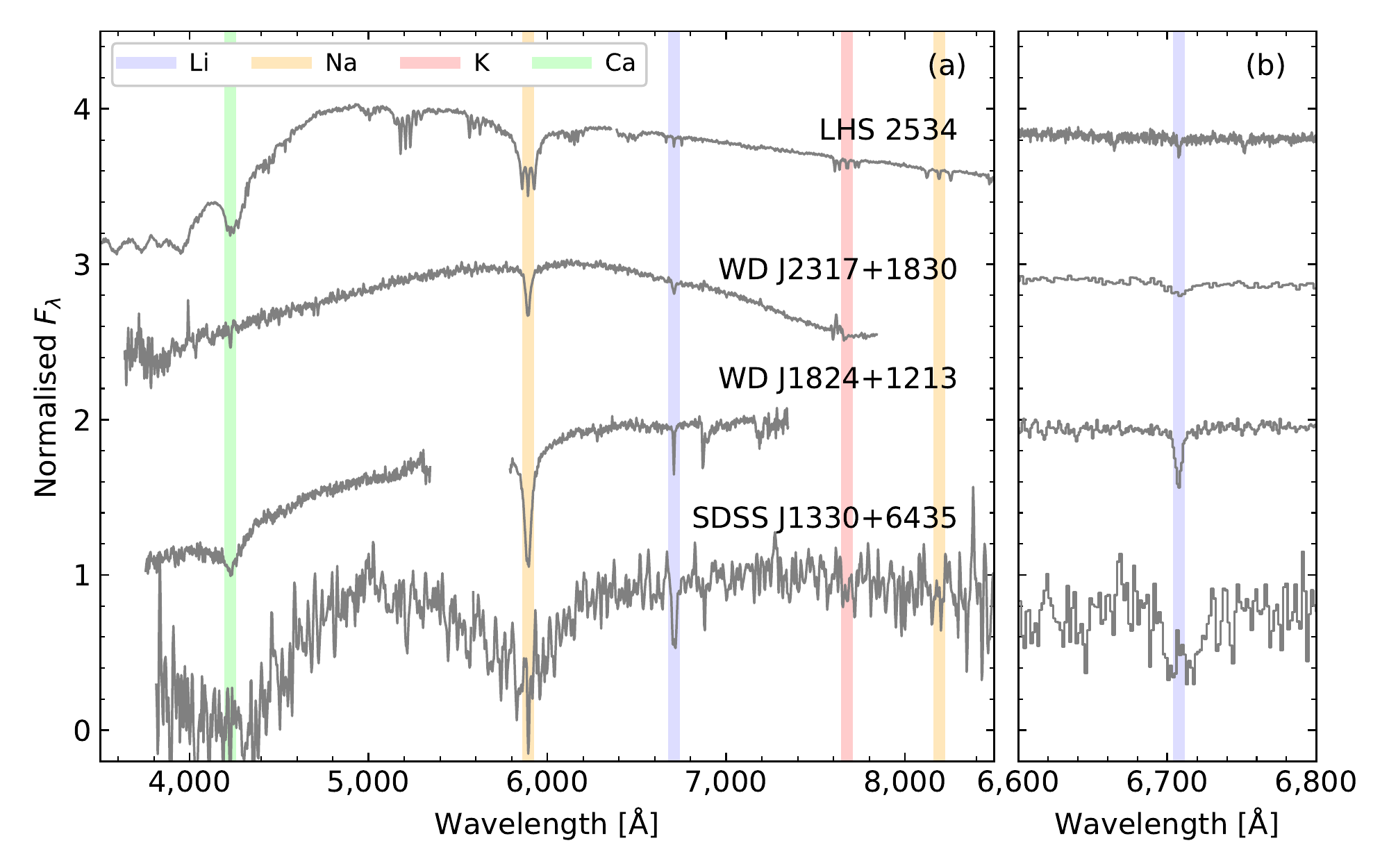}
  \caption{\label{fig:allspec}\textbf{Optical spectra of the four white dwarfs with photospheric lithium.}
  The wavelengths of the most important transitions are indicated by the colored bars. The spectra in the left-hand panel have been smoothed by a Gaussian with a full width half maximum of 3\,\AA\ for clarity, with the exception of \Lid\ where 8\,\AA\ was used instead. The spectra in the right-hand panel are not smoothed.
  }
\end{figure}

The accurate astrometry of the \textit{Gaia} mission\cite{gaiaDR2-collab-1} enabled the identification of nearby, intrinsically faint white dwarfs against much more numerous luminous background stars\cite{gentilefusilloetal19-1}, and spectroscopic observations of practically all 524 northern white dwarfs within 40\,pc are now complete\cite{tremblayetal20-1}. We have detected absorption of the lithium 6,708\,\AA\ doublet in the spectra of three cool ($\Teff<5,000$\,K) white dwarfs (Fig.\,\ref{fig:allspec}) within this sample (\Lia, \LibLong, \linebreak \LicLong), revealing the presence of this lithophile element within their photospheres.

The 6,708\,\AA\ doublet of neutral lithium is the only strong transition of this element at optical wavelengths, and because of the low ionization energy of lithium (5.4\,eV), it becomes rapidly undetectable in hotter white dwarfs. Inspecting the published spectroscopy of cool white dwarfs at distances beyond 40\,pc\cite{hollandsetal17-1, harrisetal03-1}, we identified a fourth system (\LidLong) exhibiting lithium absorption. 

Spectroscopy of planetary bodies accreted into the pristine hydrogen or helium atmospheres of white dwarfs provides direct measurements of their bulk abundances\cite{zuckermanetal07-1}, similar to the analysis of meteorites to determine the composition of solar system planets\cite{lodders03-1}. All four stars with photospheric lithium also exhibit sodium and calcium lines (Fig.\,\ref{fig:allspec}), enabling a comparative study of the volatile and refractory content of their accreted planetesimals. 
The planetesimals, or fragments thereof, are most likely scattered via gravitational interactions with more massive bodies from several au into the tidal disruption radius of the white dwarf\cite{debesetal12-1}. An alternative way of delivering planetary material to the white dwarf is the Kozai-Lidov mechanism in wide binaries\cite{petrovich+munoz17-1}, however, we do not detect wide companions for any of the four stars discussed here in Gaia DR2.

The observational data available for these objects were analyzed using a model atmosphere code that has been specifically developed to correctly treat the complex physics in the high-density atmospheres of white dwarfs\cite{koester10-1}. We fitted the effective temperature (\Teff) and the stellar radius using published broad-band photometry and parallax (see Methods and Extended Data Fig.\,\ref{tab:ast_phot}), and subsequently determined the photospheric abundances using spectroscopy (Extended Data Fig.\,\ref{tab:params}), with the procedure repeated until convergence (Extended Data Fig.\,\ref{fig:allfits}). The analysis of \Lia\ required additional effort as the star exhibits a magnetic field of 2.10\,MG (see Methods). We measured the atmospheric parameters and lithium, sodium and calcium abundances for all four stars, and also detected magnesium, potassium, chromium and iron in \Lia. The effective temperatures, $\Teff=3,350$--$4,780$\,K, are among the lowest of any debris-accreting white dwarfs\cite{hollandsetal17-1}, reflecting the selection effect imposed by the neutral lithium detection. 

\begin{figure}
  \centering
  \includegraphics[angle=0,height=0.8\textheight]{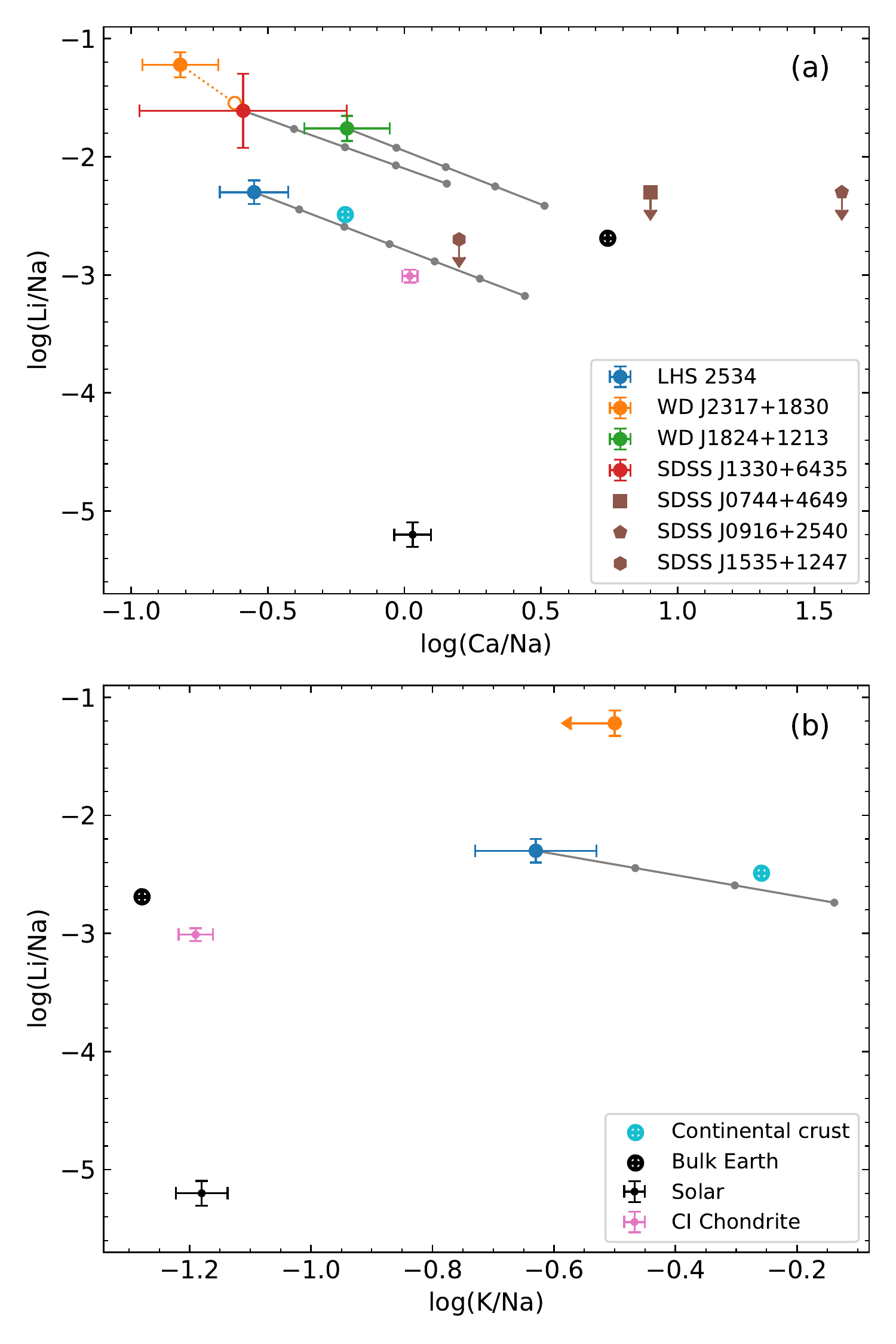}
  \caption{\label{fig:LiNaKCa}\textbf{Number density abundance ratios of debris-accreting white dwarfs and Solar system benchmarks.}
  \textbf{a}, The \logXY{Li}{Na} vs. \logXY{Ca}{Na} ratios of the four white dwarfs with lithium detections (blue, orange, green and red symbols) are enhanced with respect to the Earth's continental crust. The hollow orange point indicates the composition of the planetary body accreted by \Lib\ assuming accretion-diffusion equilibrium. The gray lines illustrate the evolution of the photospheric abundances of the other three stars due to differential diffusion if accretion has stopped in the past. The dots on these tracks indicate steps of 1\,Myr for \Lia, and 10\,Myr for \Lic\ and \Lid. Shown for comparison are solar system benchmark compositions, as well as three white dwarfs that lack a lithium detection (brown symbols). \Lia\ has also photospheric potassium, and its \logXY{Li}{Na} vs. \logXY{Ca}{Na} \textbf{a} and \logXY{Li}{Na} vs. \logXY{K}{Na} \textbf{b} ratios are consistent with having accreted a fragment of planetary crust $\simeq2$\,Myr ago. Error bars correspond to 1$\sigma$ uncertainties.
  }
\end{figure}

We compare the abundance ratios of \logXY{Li}{Na} vs. \logXY{Ca}{Na} of the four white dwarfs with those of the Sun\cite{lodders03-1}, the bulk Earth\cite{mcdonough00-1}, the continental crust\cite{rudnick+gao03-1} and CI chondrites\cite{lodders03-1} (Fig.\,\ref{fig:LiNaKCa}a). All four objects reside within a cluster, with \logXY{Ca}{Na} between $-$1 and 0, and \logXY{Li}{Na} between $-$2.5 and $-$1. Because of the rapid burning of lithium in the young Sun, the solar abundance is several orders of magnitude below those of the four white dwarfs and solar system planetary compositions. The composition of the planetary debris within the four systems is noticeably enhanced in both lithium and depleted in calcium with respect to the solar system planetary benchmarks, and most closely resembles the abundances found in the continental crust. The unusually large \logXY{Li}{Na} and low \logXY{Ca}{Na} ratios can be partially explained via differential diffusion of metals out of the convection zones since the end of the accretion episode, caused by the different elemental sinking timescales. 

We computed sinking timescales for each detected element\cite{koesteretal20-1} (see Methods, Extended Data Fig.\,\ref{tab:cvz}), and indicate the evolution of \Lia, \Lib, and \Lic due to differential evolution in Fig.\,\ref{fig:LiNaKCa}a (note the increased step sizes for \Lib\ and \linebreak \Lic). For \Lia, depending on how long ago accretion stopped, the parent body abundances could be consistent with those of the continental crust ($\approx2$\,Myr) or CI chondrites ($\approx3.5$\,Myr). At $\simeq8$\,Myr, \logXY{Ca}{Na} approaches the bulk Earth value, but \logXY{Li}{Na} would be about an order of magnitude too low. This degeneracy is broken by the additional detection of potassium in \Lia. The relative abundances of the alkali metals lithium, sodium, and potassium are entirely incompatible with either those of bulk Earth or CI chondrites, independent of the accretion history (Fig.\,\ref{fig:LiNaKCa}b). Instead, combined constraints from the lithium and potassium abundances are consistent with \Lia\ having accreted a fragment of planetary crust around 2\,Myr ago. Whereas the other three white dwarfs with photospheric lithium lack potassium detections, their close clustering near \Lia\ in \logXY{Li}{Na} vs. \logXY{Ca}{Na} strongly suggests that they, too, are contaminated by fragments of planetary crust. A similar analysis is possible for \Lic\ and \Lid, though only considering \logXY{Li}{Na} vs. \logXY{Ca}{Na} as potassium is not detected. Their past trajectories point closely towards the bulk Earth, however this origin requires many diffusion timescales to have elapsed, necessitating extremely massive parent bodies. It is therefore much more probable that smaller parent bodies were more recently accreted\cite{hollandsetal18-1} with compositions corresponding to particularly Li-rich crust. \Lib\ requires a different analysis due to its very short diffusion timescales (see below).

To investigate the compositional diversity among cool white dwarfs, we selected for comparison three additional cool white dwarfs (5,000--5,800\,K) with published high-quality optical spectra that do not show the lithium 6,708\,\AA\ line\cite{hollandsetal18-1} (\sdss{0744}{+}{4649}, \sdss{0916}{+}{2540}, \linebreak \sdss{1535}{+}{1247}, see Extended Data Fig.\,\ref{fig:morespec}). We established upper-limits for their photospheric lithium abundances, and measured calcium and sodium abundances following the same methodology as outlined above. All three stars are distinctly offset from the cluster of Li-rich white dwarfs in the \logXY{Li}{Na} vs. \logXY{Ca}{Na} diagram (Fig.\,\ref{fig:LiNaKCa}a). Their \logXY{Ca}{Na} ranges from broadly resembling CI chondrites to a value exceeding that of bulk Earth. This corroborates our hypothesis that the four white dwarfs with photospheric lithium are accreting planetary debris with a clearly distinct, crust-like, composition. 

Crust-like debris compositions have been suggested previously\cite{zuckermanetal11-1, hollandsetal18-1}, based on the detection of large calcium and aluminum abundances. However, both elements also have relatively high mass fractions within the mantle of a differentiated planet, and whereas the interpretation of calcium and aluminum-rich material as signatures of differentiation is certainly plausible\cite{bonsoretal20-1}, it does not support a definitive detection of crust fragments. In contrast, the mass fraction of the alkali metals lithium, potassium and sodium is one to two orders of magnitude higher in the Earth's continental crust than in its mantle\cite{lodders+fegley11-1}, making them unambiguous tracers of planetary crusts. 

Given that the photospheric abundances reflect the composition of the entire convection zone, in which the material remains homogeneously mixed, it is possible to estimate lower limits on the accreted planetary body masses. The properties of the outer convection zones (Extended Data Fig.\,\ref{tab:cvz}) depend sensitively on the white dwarf mass and \logXY{H}{He} ratio, and vary only mildly as function of \Teff\ within the narrow range spanned by the four stars. The masses of the individual elements contained within the convection zones are reported in Extended Data Fig.\,\ref{tab:cvz}. Scaling the sodium mass contained within the convection zone by its mass fraction of 2.4\,per cent in the continental crust of the Earth\cite{rudnick+gao03-1}, we estimate the minimum masses of the crust fragments accreted into these white dwarfs to be in the range $5\times10^{17}$--$3\times10^{20}$\,g. This compares to $\simeq4\times10^{25}$\,g for the Earth's continental crust\cite{rudnick+gao03-1}, hence the observed level of atmospheric contamination requires only relatively small splinters of the crust of an Earth-like planet. 

\begin{figure}
  \centering
  \includegraphics[angle=0,height=0.7\textheight]{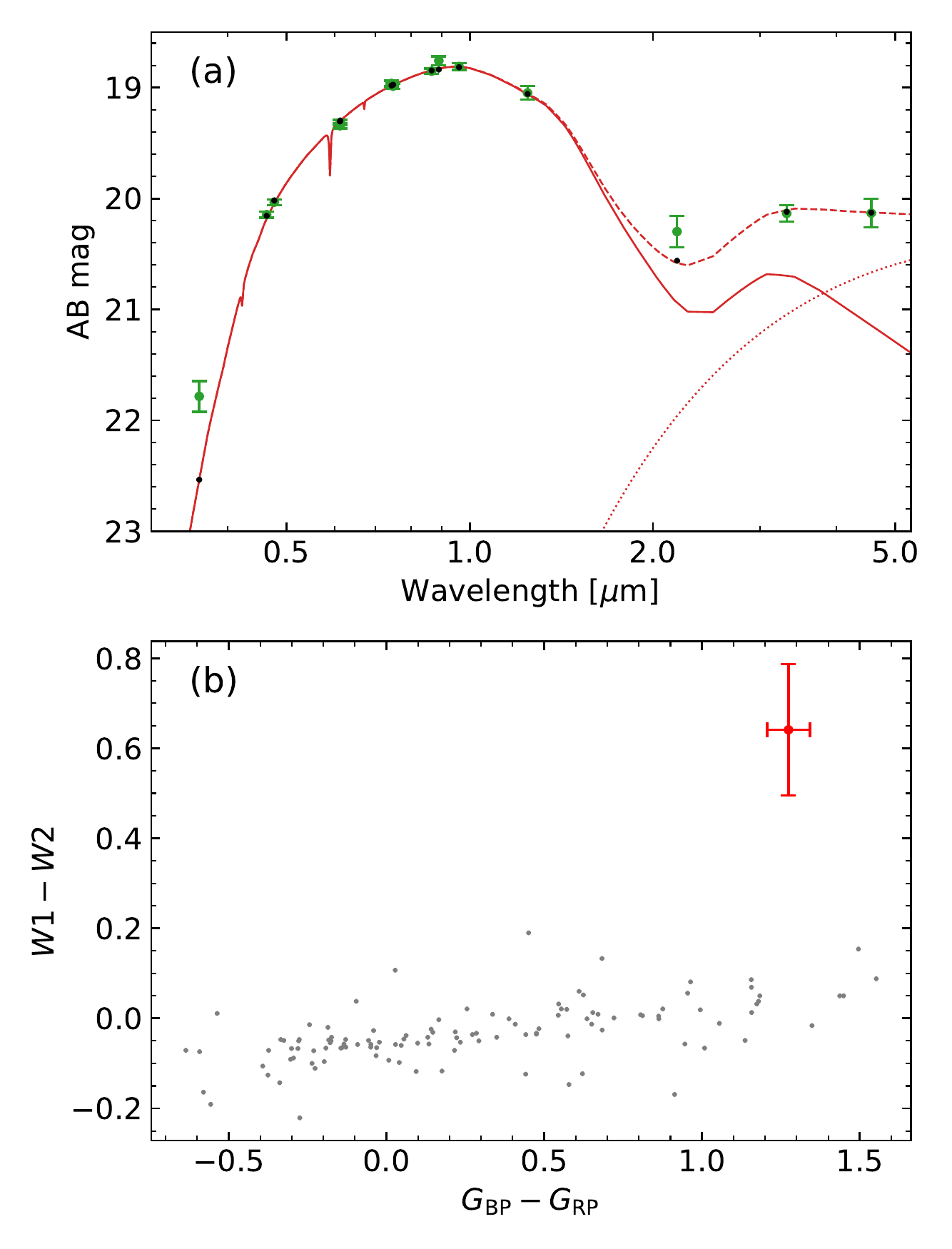}
  \caption{\label{fig:SED2317}\textbf{An infrared excess at \Lib.}
  \textbf{a}, Our best fitting white dwarf model (solid curve) for \Lib\ compared with the photometry from SDSS, Pan-STARRS, UKIRT, and \textit{WISE} shows a flux excess in the $K$, $W1$, and $W2$ bands. An opaque optically thick disk heated by the white dwarf\cite{jura03-1} with an inclination of 70\,degrees, and an inner-edge temperature of 1,500\,K (dotted line), when combined with the white dwarf flux (dashed line) provides a good fit to the photometry. \textbf{b}, \Lib\ clearly stands out with an unusually red $W1-W2$ color when compared to white dwarfs with similar $G_\mathrm{BP}-G_\mathrm{RP}$ colors (gray points, from a cross-match of \emph{Gaia} white dwarfs\cite{gentilefusilloetal19-1} with \textit{WISE}, and $W1-W2$ uncertainties $<0.05$\,mag). Error bars correspond to 1$\sigma$ uncertainties.
  }
\end{figure}

\Lib\ has a small outer convection zone, due to its large mass and mixed hydrogen/helium atmosphere (Extended Data Fig.\,\ref{tab:cvz}). The sinking timescales within \Lib\ are therefore on the order of thousands of years, much shorter than the millions of years typical of cool helium-atmosphere white dwarfs\cite{koester09-1}. Because accretion episodes are expected to last for $10^5$--$10^6$\,years\cite{girvenetal12-1} it is therefore likely that \Lib\ is still in the process of accreting from a circumstellar debris disk. This hypothesis is supported by the detection of an infrared excess (Fig.\,\ref{fig:SED2317}a), which is consistent (see Methods) with a flat, passive, optically thick disk of dust heated by the white dwarf\cite{jura03-1}. The infrared excess is detected in the UKIRT $K$ and \textit{WISE} $W1$ and $W2$ bands. For comparison, we selected 116 white dwarfs within 130\,pc from the \textit{Gaia} DR2 white dwarf catalog\cite{gentilefusilloetal19-1} and cross-matched them with \textit{WISE} detections, with the further requirement of $W1-W2$ uncertainties $<0.05$\,mag. \Lib\ is a clear outlier, presenting a 4-sigma excess in $W1-W2$ (Fig.\,\ref{fig:SED2317}b). The proper motions for \Lib\ measured from the \textit{WISE} observations obtained from 2010 to 2016\cite{eisenhardtetal20-1} agree with those determined by \textit{Gaia}\cite{gaiaDR2-collab-1}, corroborating the association of the \textit{WISE} fluxes with the white dwarf. This makes \Lib\ the coolest, and oldest white dwarf with a debris disk detection\cite{debesetal19-1}. The infrared excess of the disk is $L_\mathrm{IR}/L_\mathrm{wd}=0.06$. 

Assuming that \Lib\ is in accretion-diffusion equilibrium\cite{koester09-1}, the mass flow rates through the bottom of the convection zone are identical to the accretion rates from the debris disk, which we compute (see Methods) to be 760\,g\,s$^{-1}$, 92,000\,g\,s$^{-1}$, and 37,000\,g\,s$^{-1}$ for lithium, sodium, and calcium. Adopting that sodium accounts for 2.4\,per cent of the accreted mass (for continental crust abundances\cite{rudnick+gao03-1}) implies a total accretion rate of $\simeq3\times10^{6}$\,g\,s$^{-1}$, broadly in agreement with the expected rate driven by the low Poynting-Robertson drag exercised by this cool white dwarf on the dust\cite{rafikov11-1} and is consistent with the rates ($10^5$--$10^{11}$\,g\,s$^{-1}$) observed for other accreting systems\cite{farihi16-1}. These estimates do not account for 3D effects within convective white dwarf envelopes, and recent work\cite{cunninghametal19-1} suggests that these effects may result in higher accretion rates than found from the 1D calculations we used.

There are currently few undisputed detections of planets with host star masses \linebreak $\ge3.1$\,\Msun\cite{satoetal12-1,verasetal20-1}. Using the empirical initial-to-final mass relation\cite{cummingsetal18-1}, the high mass of \linebreak\Lib\ provides a probable indication that main-sequence stars as heavy as $4.8\pm0.2$\,\Msun, corresponding to B-type stars, form planetary systems and that they survive to white dwarf stage. Note that despite the relatively short main sequence lifetime of the progenitor, second generation planet formation can largely be ruled out, as Li-burning on the main sequence would result in Li-poor planetesimals, unlike those observed here.
\Lib\ is also one of the oldest systems known to have formed differentiated rocky planets, with a $9.5\pm0.2$\,Gyr cooling age and $9.7\pm0.2$\,Gyr total age. These mass and age measurements provide constraints for planet formation models that are extremely difficult to achieve from observations of planets around main-sequence or giant stars\cite{verasetal20-1}, and the detection of a debris disk at \Lib\ demonstrate that planetary systems, or what is left of them, can remain dynamically active for practically the age of the Galaxy.  

Lithium enrichment of giant stars has been previously interpreted by the accretion of giant planets engulfed during the red giant phase\cite{aguileragomezetal16-1} and lithium has been identified in the atmosphere of a giant exoplanet via transmission spectroscopy\cite{chenetal18-1}. The detection of lithium within white dwarf photospheres originating from the accretion of planetary crust fragments represents an important link to the overall evolution of planetary systems, providing the sensitivity to establish the composition of the crusts of differentiated rocky planets. 

\clearpage

\bibliographystyle{naturemag}
\newcounter{firstbib}

\clearpage


\begin{addendum}
\item[Author Correspondence] All correspondence regarding this work should be directed to M. Hollands.
\item
MAH and PET acknowledge useful discussions with Hans-G\"unter Ludwig and Matthias Steffen regarding neutral broadening of lithium lines. We also acknowledge Jack McCleery for maintaining a database of 40\,pc white dwarf spectra.
MAH and DK acknowledge atomic data from Thierry Leininger used for the \Ion{Ca}{i} unified profile.
This research received funding from the European Research Council under the European Union’s Horizon 2020 research and innovation programme number 677706 (WD3D).  B.T.G. was supported by the UK STFC grant ST/T000406/1.
BTG was supported by a Leverhulme Research Fellowship and the UK STFC grant ST/T000406/1 and by a Leverhulme Research Fellowship.
Funding for the Sloan Digital Sky Survey IV has been provided by the Alfred P. Sloan Foundation, the U.S. Department of Energy Office of Science, and the Participating Institutions. The SDSS web site is www.sdss.org. 
Based on observations collected at the European Organisation for Astronomical Research in the Southern Hemisphere under ESO programme 0102.C-0351. 
This article is based on observations (programme ITP08) made in the Observatorios de Canarias del IAC with the WHT operated on the island of La Palma by the Isaac Newton Group of Telescopes in the Observatorio del Roque de los Muchachos and with the Gran Telescopio Canarias (GTC), installed at the Spanish Observatorio del Roque de los Muchachos of the Instituto de Astrof{\'i}sica de Canarias, in the island of La Palma.

\item[Author contributions] 
M.A.H. performed data reduction, analysis and interpretation and wrote the majority of the text. P.-E.T. and B.T.G. contributed to the data interpretation and writing of this article. D.K. developed the model atmosphere code used for the analysis. N.P.G.F. contributed to the data reduction and analysis of photometric data. 

\clearpage
\item[Author Information] Reprints and permissions information is available at www.nature.com/reprints. Correspondence and requests for materials should be addressed to M.A.H.~(email: M.Hollands.1@warwick.ac.uk).

\item[Competing Interests] The authors declare that they have no competing interests.

\end{addendum}
\clearpage


\newpage



\clearpage

\begin{methods}

\subsection{Identification and observations.}
Three of the stars with lithium detections were found among our observations of white dwarfs within 40\,pc\cite{tremblayetal20-1} while the fourth one was identified from its SDSS spectrum\cite{harrisetal03-1}. 

\Lia\ is a nearby white dwarf located 38\,pc away, and represents the first discovered magnetic metal-contaminated white dwarf\cite{reidetal01-1}. The Zeeman effect from the 2.1\,MG magnetic field\cite{hollandsetal17-1} splits the photospheric lines into multiple components. For most transitions (where fine structure interactions are much weaker than that of the magnetic field), this results in three components separated by 98.0\,\pcm\ (46.686\,\pcm\,MG$^{-1}$ in general). On 2019 January 14 we obtained spectra of \Lia\ using X-Shooter, an intermediate resolution echelle spectrograph mounted on the Very Large Telescope (VLT) at Paranal. Two exposures of 1,250\,s each were taken in the UVB and VIS arms, with 1.0 and 0.9\,arcsec slit widths used respectively. All data were reduced using the standard procedures within the REFLEX (http://www.eso.org/sci/software/reflex/) reduction tool developed by ESO. Telluric line removal was performed on the reduced spectra using MOLECFIT\cite{smetteetal15-1,kauschetal15-1}. The X-Shooter spectra clearly reveal the \Ion{Li}{i} 6,708\,\AA\ line Zeeman-split into three components, where the depths of the $\pi$ and $\sigma$ components reach 0.14 and 0.10 of the continuum respectively.

We observed \Lib\ on 2018 September 2 using the OSIRIS spectrograph on the Gran Telescopio Canarias (GTC) during service mode observations as part of our International Time Programme (ID ITP08). We used the R1000B grating with a 1\,arcsec slit-width providing a resolving power of $\simeq 1,000$ as measured from the sky-spectrum. Debiasing, flat-fielding, and extraction of the 1D spectra were performed using packages from the Starlink collection of software. Wavelength and flux calibration were performed using Molly (http://deneb.astro.warwick.ac.uk/phsaap/software/). As the service mode flux standards were observed with a wider slit-width (2.5\,arcsec), the quality of flux calibration and telluric removal were generally poor, though this did not affect the quality of our subsequent fits, as the continuum is well defined. The two strongest features of the spectrum are the \Ion{Na}{i} 5,893\,\AA\ doublet and the \Ion{Li}{i} 6,708\,\AA\ doublet (Fig.\,\ref{fig:allspec}). The \Ion{Ca}{I} 4,227\,\AA\ line is also detected though it is much weaker.

\Lic\ was observed on 2018 August 7/8, again as part of our ITP, using the Intermediate-dispersion Spectrograph and Imaging System (ISIS), mounted on the William Herschel Telescope (WHT). We used the R600B and R600R gratings with a slit width varying between 1--1.5\,arcsec between the two nights, and employing $2\times2$ binning, resulting in an average resolution of $\approx$ 2\,\AA. The spectrum exhibits an almost saturated \Ion{Na}{i} doublet, a narrow Li 6,708\,\AA\ doublet, and a relatively broad Ca\,\Ion{i} 4,227\,\AA\ line (Fig.\,\ref{fig:allspec}).

Given the detection of lithium in three cool white dwarfs, we investigated the published spectroscopy of metal-contaminated objects with similar temperatures and identified the \Ion{Li}{i} 6,708\,\AA\ line in the SDSS spectrum of \Lid\cite{harrisetal03-1}. \Lid\ was recently revisited\cite{blouinetal19-1} where the lithium absorption feature is visible in one of the figures, though the authors did not comment on the presence of photospheric lithium. 

The detection of photospheric lithium in the SDSS spectrum of \Lid\ raised the possibility that other cool white dwarfs with lithium lines in their SDSS spectroscopy may have hitherto gone unnoticed.  We therefore carried out a search for lithium-bearing cool white dwarfs, extracting the 37,259 white dwarf candidates with SDSS spectroscopy from the \emph{Gaia}~DR2 white dwarf catalog\cite{gentilefusilloetal19-1}. We removed 7,396 objects that were classified by the authors as quasars on the base of their SDSS spectroscopy. 

We ran an automated search on the 29,863 remaining spectra, which simply put, involved fitting a Gaussian profile at the expected wavelength of the \Ion{Li}{i} 6,708\,\AA\ line, and then measuring the significance. More specifically, we clipped each spectrum to the range 6,610--6,810\,\AA, which we then normalized via a fit with a first-order polynomial, excluding the region 6,690--6,730\,\AA. At the low resolution of the SDSS spectroscopy ($R\simeq2,000$), the \Ion{Li}{i} 6,708\,\AA\ doublet is not resolved, and we therefore fitted it with a Gaussian with a width of $\sigma = 5$\,\AA, where the amplitude of the Gaussian was the only free parameter. Given the width of the Gaussian, even large radial velocity shifts of 200\,\kms\ are contained well within the fitted profile. We visually inspected all spectra where the amplitude-parameter was measured to a significance of $>5\sigma$ in the direction of absorption, and where the reduced $\chi^2$ was less than 2.0.

This process easily recovered \Lid\ with an amplitude of $7.3\sigma$ and a reduced $\chi^2 = 1.033$. After  removing multiple false positives (mostly magnetic hydrogen-atmosphere white dwarfs with $B\simeq7$\,MG where the $\sigma^+$-component coincides with \Ion{Li}{i} 6,708\,\AA), we did not identify any additional  white dwarfs with conclusive lithium absorption.

\subsection{Atmospheric analysis.}
\label{sec:atm}

The spectra and available photometry were analyzed using the Koester LTE model atmosphere code\cite{koester10-1}.  Some improvements to the code have been implemented since its use in previous publications, several of which are relevant to this work. Improvements to the equations of state have been made to accommodate the very cool nature for these stars (\Lic\ in particular). Additionally, a unified profile for the \Ion{Ca}{i} 4,227\,\AA\ resonance line broadened by neutral helium is now used. Our calculations used the potentials and dipole moments provided by Thierry Leininger and recently presented with similar profile calculations\cite{blouinetal19-2}.

For the photometric fitting we made use of a wide range of photometry (Extended Data Fig.~\ref{tab:ast_phot}). This included wherever available Pan-STARRS\cite{chambersetal16-1}, SDSS\cite{alametal15-1}, and SkyMapper\cite{wolfetal18-1} in the optical. In the infrared we used 2MASS\cite{cutrietal03-1,skrutskieetal06-1}, UKIRT\cite{valentinuzzietal09-1,morettietal14-1,dyeetal18-1}, and WISE\cite{cutrietal13-1}. Note that for the latter, we have made use of the recent catWISE catalog\cite{eisenhardtetal20-1} which provides $W1$ and $W2$ detections for all four stars discussed here. 

Prior to the photometric fit, we converted all magnitudes to the AB scale. However we chose to exclude the \textit{Gaia} photometry, because when using the provided AB zero-points, we found consistent disagreement with other optical photometric surveys, i.e. SDSS, Pan-STARRS, and Skymapper (which were always found to be mutually in agreement).  In addition to the main parameters of \Teff\ and radius, the parallax is included as a dummy parameter, with the \textit{Gaia} values serving as Gaussian priors. This has the desired effect of correctly folding the parallax uncertainty into our radius estimates, and to some extent into the \Teff\ estimates. The hydrogen and metal abundances were included in the models used in the photometric fits, though with their values fixed (to become free parameters for the spectroscopic fitting, where the \Teff\ and radius are fixed instead). The model fluxes, $F_\nu(\lambda)$, were scaled by the radius and parallax, and synthetic AB magnitudes, $m$, were calculated from
\begin{equation}
    \langle F_\nu \rangle = \frac{\int \mathrm{d}\lambda\,F_\nu(\lambda) S(\lambda)/\lambda}
    {\int \mathrm{d}\lambda\,S(\lambda) /\lambda},
\end{equation}
and
\begin{equation}
    m = -2.5\log_{10}(\langle F_\nu \rangle/3,631\,\mathrm{Jy}),
\end{equation}
where $S(\lambda)$ is the energy-counting filter response curve. The atmospheric parameters were then fitted via $\chi^2$-minimization between the observed and synthetic magnitudes. For these nearby white dwarfs, the effects of interstellar reddening can be considered negligible. An important caveat regards the photometric uncertainties, which for some of the deep surveys such as Pan-STARRS, can be as small as a few mmag. These data tend to dominate the fit, and result in unrealistically small uncertainties ($<10$\,K on \Teff). It is therefore important to note that these photometry may have a relative precision at the mmag level, in particular when derived from stacked multi-epoch observations, however, the absolute fluxes have additional systematic uncertainties. To account for this, we added a constant systematic uncertainty to all available photometry of a given object. The magnitude of this systematic uncertainty was varied until the best-fit had a reduced $\chi^2$ of one. We found values of $\simeq0.04$\,mag were required for \Lia, \Lib, and \Lic. \Lid\ is sufficiently faint (the photometric errors are sufficiently large), that this step was not necessary.

For the spectroscopic fitting, we used fixed $\Teff$ and radius values derived from the photometric fit. The surface gravity, $\log g$, was calculated from \Teff\ and the radius using the white dwarf mass-radius relation\cite{fontaineetal01-1}. For \Lia\ and \Lid\ where the atmospheric hydrogen content is trace at most, we adopted the thin hydrogen mass-radius relation, whereas for \Lib\ and \Lic\ which have mixed hydrogen/helium atmospheres, we used thick hydrogen relation instead. The flux calibration of our spectroscopy is of variable quality, therefore for \Lib\ and \Lic\ (where the continuum is well defined), at each step in the least-squares fit, we performed a local normalization of the data, fitting a spline to the ratio of the observed spectrum and the model in order to re-scale the spectral fluxes.  For \Lia\ and \Lid, we instead re-calibrated the spectral fluxes against the available optical photometry fitting the difference between observed and synthetic photometry with 2nd order polynomials.

These spectroscopic and photometric fits were then iteratively repeated until convergence was found between the two solutions. The abundance errors obtained from the covariance matrix of the spectroscopic fits only considered the statistical uncertainty around a fixed \Teff, and so were typically unrealistically too small (e.g. $<0.03$\,dex). Therefore the spectroscopic fits were repeated with the best fit \Teff\ increased by $1\sigma$, with the subsequent shift in abundances added in quadrature to the statistical uncertainties.

Additional details for each object are summarized in the following sub-sections, including any departures from the general approach described above. The final results for all four white dwarfs are compiled in Extended Data Fig.~\ref{tab:params}. The best fitting models to each of the stars are shown in Extended Data Fig.\,\ref{fig:allfits}.

\subsection{Analysis of \Lia.}
\label{sec:WDa}

As the brightest of our four objects, \Lia\ has a multitude of photometry, covering its entire spectral energy distribution (SED). However, the main challenge to fitting this object is its 2.1\,MG magnetic field. Our models are intrinsically non-magnetic, therefore to improve the accuracy of our fits within this limitation, we triplicated the majority of spectral lines, reducing the $\log gf$ values by $\log_{10}(3)$ for each component. Exceptions are the \Ion{Ca}{ii}, and \Ion{Mg}{ii} resonance lines for which we used pre-computed unified line profiles.

With the higher resolution spectroscopic data, we immediately determined that the Zeeman triplet located between 5,100--5,300\,\AA\ is in fact not from \Ion{Mg}{i}\cite{reidetal01-1}, but rather \Ion{Cr}{i} as the central component has a rest-frame (air) wavelength of 5,207\,\AA. However, we note that the $\sigma^+$ component of \Ion{Mg}{i} is possibly visible at 5,155\,\AA\ with its other components blended with \Ion{Cr}{i}. The magnesium abundance is further constrained from the red wing of the 2,852\,\AA\ \Ion{Mg}{i} resonance line. Other notable spectral features are \Ion{K}{i} lines (discussed below), Zeeman split lines from the \Ion{Na}{i} 8,191\,\AA\ doublet, and a complex splitting pattern from the \Ion{Ca}{ii} 8,600\,\AA\ triplet.

As well as lithium, for \Lia\ we also detected potassium from Zeeman split \Ion{K}{i} lines. We did not detect this element for any of the other objects, however we note that for \Lic\ our wavelength coverage does not extend red enough, and for \Lid\ the spectrum is of too poor quality to infer the presence of \Ion{K}{i} lines. For \Lib\ we were instead able to obtain an upper-limit (see below). Therefore future observations may also reveal potassium in these stars. Confirming the six components observed around 7,680\,\AA\ (Supplementary Fig.\,\ref{fig:KI}) did indeed belong to \Ion{K}{i} was complicated by the fact that the interaction of the 2.1\,MG magnetic field is comparable to the fine-structure energy separation of the \Ion{K}{i} doublet. For much smaller fields the doublet components will be split into four and six sub-components, according to the anomalous Zeeman effect. In higher fields where the spin-orbit interaction is disrupted, three components separated by $\mu_BB$ would be observed instead. 2.10\,MG falls into the intermediate regime, instead observing a triplet of doublets (Supplementary Fig.\,\ref{fig:KI}). 
To determine the wavelengths empirically, the energies of the upper levels can be found from the eigenvalues of
\begin{equation}
  \small
  \left(\begin{matrix}
    E_1+2\beta & 0 & 0 & 0 & 0 & 0 \\
    0 & E_1+\tfrac{2}{3}\beta & 0 & 0 & \frac{\sqrt{2}}{3}\beta & 0 \\
    0 & 0 & E_1-\tfrac{2}{3}\beta & 0 & 0 & \frac{\sqrt{2}}{3}\beta \\
    0 & 0 & 0 & E_1-2\beta & 0 & 0 \\
    0 & \frac{\sqrt{2}}{3}\beta & 0 & 0 & E_2+\tfrac{1}{3}\beta & 0 \\
    0 & 0 & \frac{\sqrt{2}}{3}\beta & 0 & 0 & E_2-\tfrac{1}{3}\beta \\
  \end{matrix}\right),
\normalsize
\end{equation}
where $E_1 = 13,042.90$\,\pcm\ and $E_2=12,985.19$\,\pcm\ (the energies of the unperturbed $^2P_{3/2}$ and $^2P_{1/2}$ respectively), and $\beta=46.686\,\pcm\mbox{MG$^{-1}$}\times B$. Because the lower $^2S_{1/2}$ level (the potassium ground state) has no orbital angular momentum, the perturbed energies of the two sub-states are simply $\pm\beta$. Valid transitions can be found according to the normal selection rules, i.e. $\Delta m_J = 0, \pm1$ and $\Delta m_S = 0$, where $m_J$ and $m_S$ have their usual meanings. Setting $B$ to 2.10\,MG, we determined the wavelengths of the six allowed transitions finding excellent agreement with the observed data (Supplementary Fig.\,\ref{fig:KI}), thus confirming the transitions belong to potassium.

We started our fits by adopting earlier published parameters\cite{hollandsetal17-1}. Within the least squares fit to the spectrum, we found that we required a moderate hydrogen abundance, allowing for a better match to the continuum slope redward of 5,000\,\AA, which was too steep for our initial hydrogen abundance of $\logXY{H}{He}=-5$, and also a better match to the flux level blueward of 4,000\,\AA. Some of the metal lines in our the spectrum are much narrower than we were able to reproduce in our models (the cores of the \Ion{Na}{i} 5,892\,\AA\ and the \Ion{Na}{I} 8,191\,\AA\ doublet, the lithium and potassium lines, the \Ion{Cr}{i} lines, and the many Zeeman-split components of the \Ion{Ca}{ii} 8,600\,\AA\ triplet), unless the hydrogen abundance was raised to $\logXY{H}{He}\simeq-1$, though this came at the expense of substantially reducing the fit-quality to other spectral features, in particular the \Ion{Ca}{i} 4,227\,\AA\ resonance line. Because the measurement of lithium and potassium abundances are crucial to our discussion, we instead optimized those elements to match the equivalent width. We discuss the narrowness of lithium lines in particular below.

From our fit to the photometry we derived a \Teff\ somewhat cooler than found in previous  analyses\cite{reidetal01-1,hollandsetal15-1,hollandsetal17-1}. However, those studies were performed before the release of \emph{Gaia} DR2 astrometry, which places an additional constraint on the radius (and therefore mass and $\log g$). We note that at the higher temperature of 5,200\,K found in a previous analysis\cite{hollandsetal17-1}, we are forced to consider a $\log g$ of 8.22, whereas with our \Teff\ of $4,780\pm50$\,K requires a more typical surface gravity of $7.97$. Furthermore, at the higher temperature we also required a hydrogen abundance closer to $\logXY{H}{He}=-1$, in order to suppress the level of flux bluewards of $4,000$\,\AA, though as noted above, such a hydrogen abundance causes problems fitting other features of the spectrum. Because we computed non-magnetic models, as with all cool metal-contaminated  white dwarfs, we find convection up to the photosphere. However, it has been found that magnetic fields can inhibit convective processes\cite{gentileetal18-1}, which is likely the case for \Lia\ since we derive a plasma $\beta$-parameter value of $\approx$ 0.5 in the photosphere\cite{tremblayetal15-3}, indicating that magnetic pressure dominates over thermal pressure. However calculating our models without convection results in only minor changes to the emergent spectrum. Furthermore we emphasize that with such marginally low plasma $\beta$ value, convective velocities are damped by less than one order of magnitude. These convective velocities are still large compared to microscopic diffusion velocities\cite{cunninghametal19-1}, suggesting that metals are still efficiently mixed, similarly to the non-magnetic case.

\subsection{Analysis of \Lib.}

For this white dwarf photometry is available from SDSS, Pan-STARRS, UKIRT, and \textit{WISE}. The UKIRT $J$ magnitude is from the UKIRT Hemisphere Survey (UHS)\cite{dyeetal18-1}, with the $K$ band from the Wide-field nearby Galaxy-cluster survey (WINGS)\cite{valentinuzzietal09-1,morettietal14-1}. The SDSS $u$-band is poorly fitted in Extended Data Fig.\,\ref{fig:allfits}, though the detection flags indicate that the uncertainty is underestimated in this filter\cite{alametal15-1}. The combination of UKIRT and \textit{WISE} photometry reveal collision induced absorption (CIA) is present in the atmosphere of this star.

We found that it was not possible to fit all the near-infrared photometry simultaneously. Instead we used only the optical and $J$-band photometry to constrain the \Teff\ and radius. This produced a much more consistent fit with spectroscopy, where the width of the Na doublet and $y-J$ color are both sensitive to the He/H ratio. Evidently from Extended Data Fig.\,\ref{fig:allfits} this results in too little flux in the $K$ and \textit{WISE} bands. We noted that the $W1-W2$ color is far too flat compared to the expected Rayleigh-Jeans
tail. For two photometric measurements along a Rayleigh-Jeans tail, it is expected that
\begin{equation}
    \frac{m_2-m_1}{\log_{10}(\lambda_2/\lambda_1)}\simeq5,
    \label{eq:maggrad}
\end{equation}
where $m_{1,2}$ are the two AB magnitudes, and $\lambda_{1,2}$, their corresponding central wavelengths. For example, for the $W1/2$ photometry of \Lia, equation~\eqref{eq:maggrad} is evaluated to be $4.7\pm0.8$, whereas for \Lib\ the same quantity is $0.0\pm1.1$~--~more than 4$\sigma$ smaller than the expected value for a Rayleigh-Jeans slope. We highlight this excess
in Fig.\,\ref{fig:SED2317}b where we show the $W1-W2$ color against the $G_\mathrm{BP}-G_\mathrm{RP}$ color for \Lib\ and a
cross-match of the \emph{Gaia} white dwarf catalog\cite{gentilefusilloetal19-1} and \emph{WISE} photometry\cite{wrightetal10-1,mainzeretal11-1}. Initially this cross-match contained 70,050 sources. We further refined
this be keeping only sources with white dwarf probabilities, $P_\mathrm{wd} > 0.75$, reducing the sample to 28,333. From a color-color diagram of $G_\mathrm{RP}-W1$ vs. $G_\mathrm{BP}-G_\mathrm{RP}$ it was clear that the cross match was contaminated by a larger number of sources with very red $G_\mathrm{RP}-W1$ (i.e. flux-contamination from nearby sources, white dwarfs with main sequence companions). We therefore made a cut of $G_\mathrm{RP}-W1 > 0.2 + 1.3(G_\mathrm{BP}-G_\mathrm{RP})$, leaving only the main white dwarf locus, containing 4,076 objects. Finally, we removed objects with $W1-W2$ uncertainties $>0.05$\,mag, leaving only the 116 objects shown in Fig.\,\ref{fig:SED2317}b, all of which are contained within 130\,pc.

We considered the possibility that the perceived flux excess of \textit{WISE} photometry result from contamination by another object located within the large \textit{WISE} PSF. However, the catWISE\cite{eisenhardtetal20-1} photometry collected over the period 2010 to 2016 shows that the source detected by \textit{WISE} has a proper motion consistent with that measured by \textit{Gaia}, strongly arguing against background contamination. Furthermore the excess is also seen in the UKIRT $K$-band, which has a much better spatial resolution.

An possible explanation that could resolve the discrepancy between the observed SED and the model spectra would be a modification of the CIA absorption profiles, as they have been subject to limited observational tests. While simply increasing the strength of the absorption would extend the red-wing of the H$_2$-He opacity into the $W1$ band, this would also result in excess absorption in the $K$-band. However, the H$_2$-He CIA opacity has been shown\cite{blouinetal17-1} to be distorted at the high densities relevant to these stars ($>0.1$\,\gcc). The opacity table used in this work has only temperature dependence and so it is feasible that density-dependent shifts may explain our observations. Even so, a strong argument against this, comes from the fact that we were able to fit \Lic\ accurately in all photometric passbands, despite this star having stronger CIA due to its lower \Teff.

A more natural explanation is that the infrared excess arises from a circumstellar dust disk irradiated by the white dwarf. Using the simple flat disk model\cite{jura03-1}, we found that a reasonable fit is obtained for an inner disk-temperature of $\simeq1,500$\,K outer temperature of $>500$\,K and an inclination of $\simeq70$\,degrees (dashed curves in Fig.\,\ref{fig:SED2317}a and Extended Data Fig.\,\ref{fig:allfits}). The fact that \Lib\ has short diffusion timescales, which make it likely that we observe the star actively accreting corroborates the detection of an infrared as the signature of a dusty debris disk, formed from the tidal disruption of a planetesimal.

In addition to \Lia, the spectrum of \Lib\ also covers the wavelength region of the potassium doublet.
Strong telluric absorption coincides with the wavelength of the blue component,
making it impossible to extract a meaningful upper-limit from this line -- even with telluric removal (as in Fig.\ref{fig:allspec}), large residuals remain.
The red component resides outside the region covered by telluric absorption, and so we were able to use
this line to establish an upper-limit of $\logXY{K}{He}<-10.5$. This results in an abundance ratio upper-limit of
$\logXY{K}{Na}<-0.5$. Since this is higher than the measurement of \Lia\ (Fig.\,\ref{fig:LiNaKCa}b),
the true value for \Lib\ could be several 0.1\,dex below the upper-limit and be broadly consistent with the location
of \Lia\ in Fig.\,\ref{fig:LiNaKCa}b.

\subsection{Analysis of \Lic.}
\label{sec:WDc}

For this object, optical photometry are available from Pan-STARRS, though because of the $1,200$\,\masy\ proper-motion, it appears as four separate detections, where we list the weighted averages provided in Extended Data Fig.~\ref{tab:ast_phot}. Additionally, \Lic\ has a UKIRT $J$-band detection in the UHS\cite{dyeetal18-1}. \Lic\ was found to be an ultra-cool white dwarf ($\Teff=3,350\pm50$\,K) with a mixed H/He atmosphere similar to \Lib. This results in very strong CIA from H$_2$-He, to the extent that its effect is measurable from only the Pan-STARRS $z-y$ color, though it is further constrained by the $J$-band. $W1$ and $W2$ magnitudes were also found within the catWISE data, and agree well with our best fitting model, despite the very strong CIA. The most notable features in the spectral data are the near-saturated sodium and calcium lines, the latter of which shows a reasonable fit with our unified profile.

\subsection{Analysis of \Lid.}
\label{sec:WDd}

This stellar remnant has been recently modeled\cite{blouinetal19-1} owing to its extremely strong and broad sodium absorption. While the the unified sodium line profile used in that analysis is not currently implemented in our models, we were able to obtain an adequate fit to the data.

Using the SDSS and Pan-STARRS photometry we found a \Teff\ and $\log g$ lower than that derived in the earlier analysis\cite{blouinetal19-1}. Regardless, our model synthetic photometry is mostly in good agreement with the data, though we note that our model slightly under-predicts the flux in the $W1$ band compared with the measurement.

Two spectra available from SDSS, though the second spectrum (taken with the BOSS instrument) is of very poor signal-to-noise ratio. Even so, we decided to co-add these spectra to improve the spectroscopic uncertainties. From comparison with the photometry (Extended Data Fig.~\ref{tab:ast_phot}), the flux calibration of both spectra was found to be poor, therefore re-calibration against the photometry (as described above) was required for each spectrum before they could be co-added. We found co-adding provided a modest S/N boost of 10\,per cent in the red, and 20\,per cent in the blue (relative to the better of the two spectra).

The main challenge in fitting this spectrum was measuring the calcium abundance. The calcium resonance line is modeled here using the aforementioned unified profile and yields a reasonable fit in Extended Data Fig.\,\ref{fig:allfits}. However, we found that the calcium and magnesium abundances were highly anti-correlated~--~because the calcium resonance line is essentially saturated even at a low abundance, a similarly good fit to both the spectrum and $u$-band photometry can be achieved by lowering the calcium abundance and increasing the magnesium abundance. There is insufficient information that could constrain the Mg/Ca ratio, and so a higher quality spectrum will be needed to measure these abundances independently. Therefore our adopted calcium abundance (Extended Data Fig.~\ref{tab:params}) assumes for $\logXY{Mg}{Ca} = 1.5$\,dex, which is the average value found in an earlier study\cite{blouin20-1}.

We note that the Na and Ca abundances that we determined are significantly lower than those in the previous analysis\cite{blouinetal19-1}. This difference arises from our lower \Teff\ measurement, resulting in a changed excitation balance. Therefore we required much lower Na and Ca abundances to produce the same strength lines. Even so, we find our Ca/Na ratio is consistent with the previous analysis ($\logXY{Ca}{Na}=-0.6\pm0.4$ vs. $-0.3\pm0.4$ in the previous work). The remaining difference in the Ca/Na ratio (while already within 1$\sigma$ uncertainty) could be attributed to the anti-correlation between Ca and Mg as described above.

Given the highly pressure broadened Na doublet, it is clear that the atmosphere of \Lid\ must be helium-dominated, though trace hydrogen could be present at some level. Our best fit was performed assuming $\logXY{H}{He}=-6$ dex, though we were still able to establish an upper-limit. We found that up to $-4$\,dex, the models looked close to identical, with a small amount of CIA appearing at $2$\,$\mu$m, but otherwise continuing to agree well with all available photometry, and spectroscopy. However at $-3$\,dex we found that CIA pushed down the infrared flux compared to the optical enough to be in disagreement with the WISE photometry. Furthermore, the Na doublet started to become narrower than observed in the spectrum. Therefore we adopt $-4$\,dex as our upper-limit for the hydrogen abundance.

\subsection{Line widths of Lithium.}
\label{sec:Liwidths}

In our initial attempts to derive abundances we found that the lithium lines were systematically much narrower than predicted by our models. On its own, these narrow lines might suggest an interstellar origin\cite{ferlet+dennefeld84-1}, however there are multiple arguments against this. Firstly, for \Lia\ the lithium doublet is Zeeman split by the 2.1\,MG magnetic field. Secondly all objects apart from \Lid\ are located within 40\,pc of the Sun (\Lid\ is located $87\pm7$\,pc away) where interstellar absorption can be considered negligible. Finally, in all four objects the observed lithium lines~--~while much narrower than those in the models~--~are still broader than the instrumental resolution. We determined this by first measuring the spectral resolution at the location of the lithium-doublet by determining the widths of sky emission lines in the sky spectrum. In each object, we then fitted the lithium-doublet with a Voigt profile, with the Gaussian-component, $\sigma$, fixed to the spectral resolution, and with the Lorentzian-component, $\gamma$, as a free-parameter. The results are given in Supplementary Table~\ref{tab:Liwidths}, and include the radial velocities measured from the Voigt profiles (not corrected for gravitational redshift). We thus conclude that in all four cases the lines must be photospheric in origin.

Instead we considered the possibility that the overly broad lines in our models arise from inaccurate atomic data (uncertainties can often be as large as a factor 2--3 (0.3--0.5\,dex). For lithium, we obtained atomic data from VALD\cite{ryabchikovaetal97-1,ryabchikova15-1}, where the broadening constants $\log(\Gamma_\mathrm{rad})$, $\log(\Gamma_\mathrm{stark}/n_\mathrm{e})$, $\log(\Gamma_\mathrm{VdW}/n_\mathrm{H})$ were found to be $7.56$, $-5.78$, and $-7.57$ respectively. These broadening constants are nominally calculated for temperatures of 10,000\,K and for a single perturber, and so internal scaling is required for different temperatures, densities, and in the case of neutral broadening, other perturbers such as He. In the cool, dense atmospheres of these four white dwarfs, the dominant line-broadening process is from perturbations by neutral particles, i.e. hydrogen, helium, H$_2$. Indeed we found that the lithium line widths in all objects are sensitive to adjustments in the neutral-broadening constant. We therefore decided to empirically determine a correction to the neutral-broadening constant, $\log(\Gamma_\mathrm{VdW}/n_\mathrm{H})$, from our observations.

We decided to perform this measurement on a single object, comparing the other three stars for consistency.
We chose \Lic\ as the calibration object because its $\gamma$-measurement has the highest
relative precision (Supplementary Table~\ref{tab:Liwidths}). Furthermore \Lic\ has the second highest
$\gamma/\sigma$ ratio~--~$\gamma/\sigma$ is higher for \Lid, but the low signal-to-noise
ratio and other challenges in fitting this object make it a poor
choice of calibrator.

We generated a grid of models with $\log(\Gamma_\mathrm{VdW}/n_\mathrm{H})$, spanning $-9.0$ to $-8.0$ in
steps of $0.1$ dex, and \logXY{Li}{H} spanning $-12.50$ to $-11.00$ in 0.25\,dex steps.
Other atmospheric parameters (\Teff, $\log g$, other abundances), were set to the best fitting values in
Extended Data Fig.\,\ref{tab:params}.
The optimal broadening constant was determined via a non-linear least-squares fit to the data, interpolating
the models (convolved by the 1.9\,\AA\ instrumental broadening)
at arbitrary abundance/$\log(\Gamma_\mathrm{VdW}/n_\mathrm{H})$.
We found the best fitting value to be $\log(\Gamma_\mathrm{VdW}/n_\mathrm{H}) = -8.53\pm0.06$,
or in other words a change of $-0.96\pm0.06$\,dex.

The best fitting models for all four white dwarfs with the revised broadening constant are displayed in red in Supplementary Fig.\,\ref{fig:Li_lines}. For comparison, models with the original broadening constant are shown in orange, although with abundances revised upwards by 0.3\,dex for clarity. Naturally the improvement for \Lic\ is exemplary, given that this object was used for calibration, though the line width for \Lib\ (which has a similar derived atmospheric H/He ratio) was also found to show good agreement when using the revised values. \Lia\ and \Lid, which both have He-dominated atmospheres, show improvement compared to the original broadening constant (where the lithium lines are completely washed out), though the lines remain far wider than observed in the data. The remaining discrepancy is particularly pronounced for \Lia, though other lines in the spectrum of this object are also found to be narrower than expected. This includes the potassium lines (Supplementary Fig.\,\ref{fig:KI}), and 5,892 and 8,191\,\AA\ doublets of sodium. The \Ion{Na}{i} 5,892\,\AA\ lines are particularly noteworthy, because we were able to accurately fit their wings (Extended Data Fig.\,\ref{fig:allfits}), while narrow cores remain which we are unable to fit with our models. Since we have no trouble fitting these sodium lines in the other three objects, we take this as evidence that the 2.1\,MG magnetic field has a strong effect on the atmospheric structure, exacerbating the narrowness of weaker lines. Indeed we find that the gas pressure in our models is $10^{10.5}$\,dyn\,cm$^{-2}$, whereas the magnetic pressure must be $10^{11.2}$\,dyn\,cm$^{-2}$ given the 2.10\,MG magnetic field. Therefore, in addition to only adjusting the lithium lines, for this object, it was also necessary to reduce the neutral broadening constants for \Ion{Cr}{i} and \Ion{K}{i} by the same amount, to better estimate the photospheric abundances.

While we clearly find improvement with the revised broadening constant, 
similar issues are not encountered in other cool stellar atmospheres
with lithium lines, for instance the Solar atmosphere and those of giant stars.
However calculations of giant star atmosphere models with the reduced neutral broadening constant show no discernible difference in the line width (M. Steffen, private communication, June 2020),
indicating that other broadening processes dominate within those atmospheres.

\subsection{White dwarf masses and evolution.} 

For two of the stars in our sample, \Lic\ and \Lid, we derive particularly small masses ($0.28\pm0.03$ and $0.38\pm0.06$\,\Msun, respectively). These masses are presumed to be unrealistic, as the Galaxy is too young to produce such low-mass white dwarfs, and are representative of the challenges common to modeling cool white dwarfs\cite{mccleeryetal20-1} with $\Teff<5,000$\,K. In fact, extremely low masses are commonly derived for white dwarfs with strong CIA absorption\cite{kilicetal20-1} and imply missing opacity in the stellar models. Such difficulties are often understood to arise from strongly wavelength-dependent opacities such as CIA\cite{blouinetal17-1} and the red wing of Lyman $\alpha$\cite{kowalski+saumon06-1}. 

Our models include CIA opacities\cite{karmanetal19-1,borysowetal97-1,borysowetal01-1,kowalski14-1} from H$_2$-H, H$_2$-He, H$_2$-H$_2$, H-He, and He-He-He. We also include broadening\cite{rohrmannetal11-1} of Ly$\alpha$ by H$_2$. However we do not include the effects of pressure distortion in the H$_2$-He opacity\cite{blouinetal17-1}. While we do not have access to these specific data, their inclusion in our models may go some of the way to explain these low masses. The expectation is therefore that these stars have higher true \Teff\ than from our analysis, which would therefore allow for smaller radii, needed to remain consistent with the photometry and parallaxes, which via the mass-radius relation for white dwarfs, implies higher masses. 

In the following we assume that the derived luminosities (Extended Data Fig.~\ref{tab:params}) of \linebreak \Lic\ and \Lid\ are correct  since they provide an adequate fit to the photometric SED, and additionally we assume that the mass is drawn from the distribution found for warmer white dwarfs\cite{tremblayetal13-1}, i.e. $M=0.614\pm0.122$\,\Msun. For \Lic\ we obtain $\Teff=4,050_{-240}^{+260}$\,K and  $\tau=9.41_{-0.95}^{+0.35}$\,Gyr for thick hydrogen layers which are appropriate for the large total hydrogen mass in the star. For \Lid, again assuming $M=0.614\pm0.122\,$\Msun\ leads to $\Teff=4,160_{-270}^{+290}$\,K and $\tau=7.54_{-0.60}^{+0.31}$\,Gyr for thin hydrogen layers. Clearly both white dwarfs have long cooling times but it is not possible to estimate the main-sequence lifetime owing to the uncertainty on the mass.

For \Lic\ (the most extreme case of the low-mass/low-\Teff\ systems), we refitted the photometry and spectrum with the \Teff\ fixed to 4,050\,K (as described above), and with the radius and atmospheric abundances as free parameters, to see if this may still provide an adequate solution, which could be the case if our low-mass solution was simply a local minimum. While the best model for this restricted fit did (by design) result in a mass close to 0.6\,Msun, we found the model failed to accurately reproduce the shape of the SED in the both the optical and infrared, showing particular disagreement in the $J$-band of around 0.5\,mag. Therefore we rule out the a second minimum in the parameter space at higher \Teff, though this does not discount the possibility that improved atmospheric models may shift the best solution to higher temperatures and thus towards more reasonable masses. Even so, the consistency in the abundance ratios of \Lic\ and \Lid\ compared with \Lia\ and \Lib\ (Fig.\,\ref{fig:LiNaKCa}), indicates that our conclusions on a crust origin for the accreted material is unaffected in these two low mass systems. We note that forcing an increased \Teff\ shifts the optimal abundances: \logXY{H}{He} was reduced by about $0.3$\,dex with all metal abundances increased by $0.5$--$0.6$\,dex. The similarity in metal abundance shifts
implies that the location of \Lic\ in Fig.\,\ref{fig:LiNaKCa}a, is largely unaffected by systematic uncertainty in the \Teff.

For \Lia\ we derive a mass of $0.55\pm0.02$\,\Msun\ corresponding cooling age of $5.8\pm0.2$\,Gyr. We expect the magnetic field to have negligible influence on cooling age\cite{tremblayetal15-3}. Such a white dwarf mass implies a very long main-sequence lifetime, possibly longer than the cooling time\cite{marigoetal20-1}. It is possible that \Lia\ is also impacted by the model systematics mentioned above, and therefore we refrain from estimating a total age.

In contrast, our fit to \Lib\ reveals a mass of $1.00\pm0.02$\,\Msun\ and cooling age of $9.5\pm0.2$\,Gyr making it the among most massive white dwarfs detected with signatures of a planetary system\cite{verasetal20-1}. The large mass implies a massive progenitor with a relatively small main-sequence lifetime, leading to a precise total age of $9.7 \pm 0.2$\,Gyr using an empirical initial-to-final mass relation\cite{cummingsetal18-1} and main-sequence lifetimes\cite{hurleyetal00-1}.

\subsection{Kinematics and population membership.} 
While we could only establish a reliable total age for one of the four analysed white dwarfs, kinematics can be useful to identify population \linebreak membership\cite{oppenheimeretal01-1,seabroke07-1,kilicetal19-1}. In Supplementary Table \ref{tab:kinematics}, we rely on the precise \textit{Gaia} astrometry to derive tangential velocities as well as motion in Galactic coordinates $U$, $V$ and $W$. We had to assume zero radial velocity as this quantity is poorly constrained from our spectroscopic observations. In earlier studies of halo white dwarf candidates\cite{oppenheimeretal01-1}, a $2\sigma$ halo membership required $\lvert U \rvert > 94$\,km\,s$^{-1}$ or $V > 60$\,km\,s$^{-1}$ or $V < -130$\,km\,s$^{-1}$. Using the same requirements, only \Lic\ is a halo white dwarf candidate, as previously identified\cite{kilicetal19-1}, although all our objects have relatively large tangential velocities suggesting an old disk population\cite{fantinetal19-1}, consistent with the large cooling ages.

The chemical abundances in these old white dwarfs have the potential to provide constraints on planet formation around stars formed in the early history of the Galaxy, and hence possibly under metal-poor conditions. However, early disk membership is not necessarily linked to progenitors of significant sub-solar metallicity\cite{feltzingetal01-1, casagrandeetal01-1}, and further insight will require a larger sample of cool debris-accreting white dwarfs. 

\subsection{Sinking times.} We used our new envelope code\cite{koesteretal20-1} to determine convection zone sizes and sinking timescales\cite{koester09-1} for each element considered in our sample. With only four objects we were able to use the best fit atmospheric models discussed in the previous sections as boundary conditions on the upper-envelope for self-consistency (as opposed to interpolating a grid of models). These results are listed in Extended Data Fig.~\ref{tab:cvz}. Using these timescales it is possible to trace back the atmospheric abundance histories of a metal Z (with sinking timescale $\tau$) using,
\begin{equation}
    \logXY{Z}{He}(t) = \logXY{Z}{He}(0) + \frac{t}{\ln(10)\tau},
\end{equation}
implying the relative abundances for two elements evolves as
\begin{equation}
    \logXY{Z_1}{Z_2}(t) = \logXY{Z_1}{Z_2}(0) + \frac{t}{\ln(10)}\left[\tau_1^{-1} - \tau_2^{-1}\right].
\end{equation}

The convection zone masses, combined with our abundance measurements (Extended Data Fig.~\ref{tab:params}) allow us to determine the mass of each element mixed within the convection zones, providing lower limits on the amounts of accreted material (Extended Data Fig.~\ref{tab:cvz}). In the case of \Lib, where accretion-diffusion equilibrium has been
assumed, the elemental diffusion fluxes (equal to the elemental accretion rates) can be calculated by dividing
the respective convection zone masses by their corresponding diffusion timescales.

\end{methods}


\clearpage
\begin{addendum}
\item[Data Availability] The data that support the plots within this paper and other findings of this study are available from the ESO science archive facility, the GTC public archive, ING archive, and SDSS database; or from the corresponding author upon reasonable request.

\item[Code Availability] The Koester model atmosphere and envelope codes are not publicly available,
though details of their functionality can be consulted from the included references.
\end{addendum}

\clearpage


\begin{efigure}
  \centering
  \includegraphics[angle=0, scale=0.29]{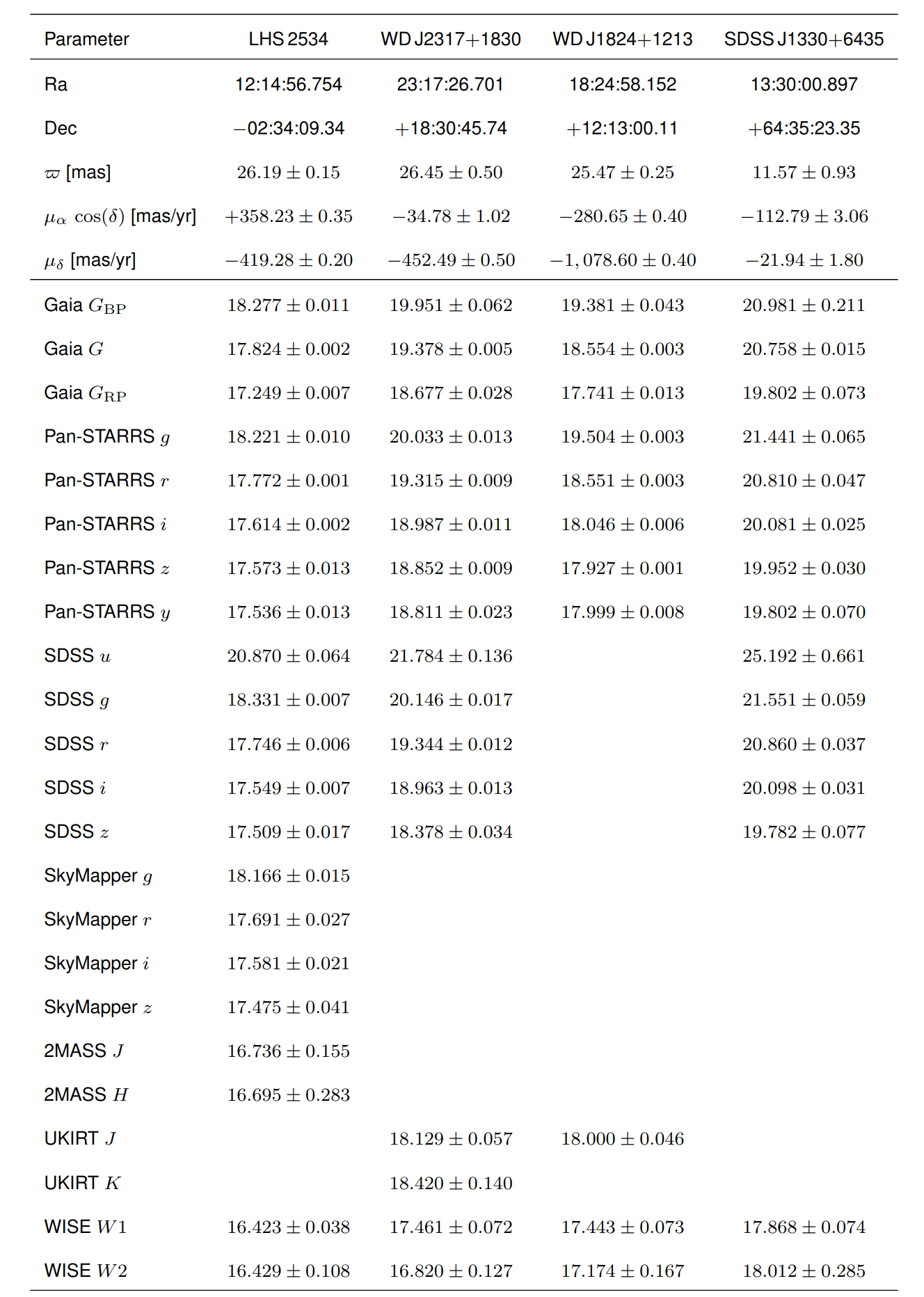}
  \caption{\label{tab:ast_phot}\textbf{Astrometry and photometry for the four
  Li-rich white dwarfs.} Pan-STARRS, SDSS and SkyMapper photometry are given
  in the AB-system, with the remainder in the Vega-system. Positions are given in the J2015.5 epoch.
  }
\end{efigure}

\begin{efigure}
  \centering
  \includegraphics[angle=0, scale=0.29]{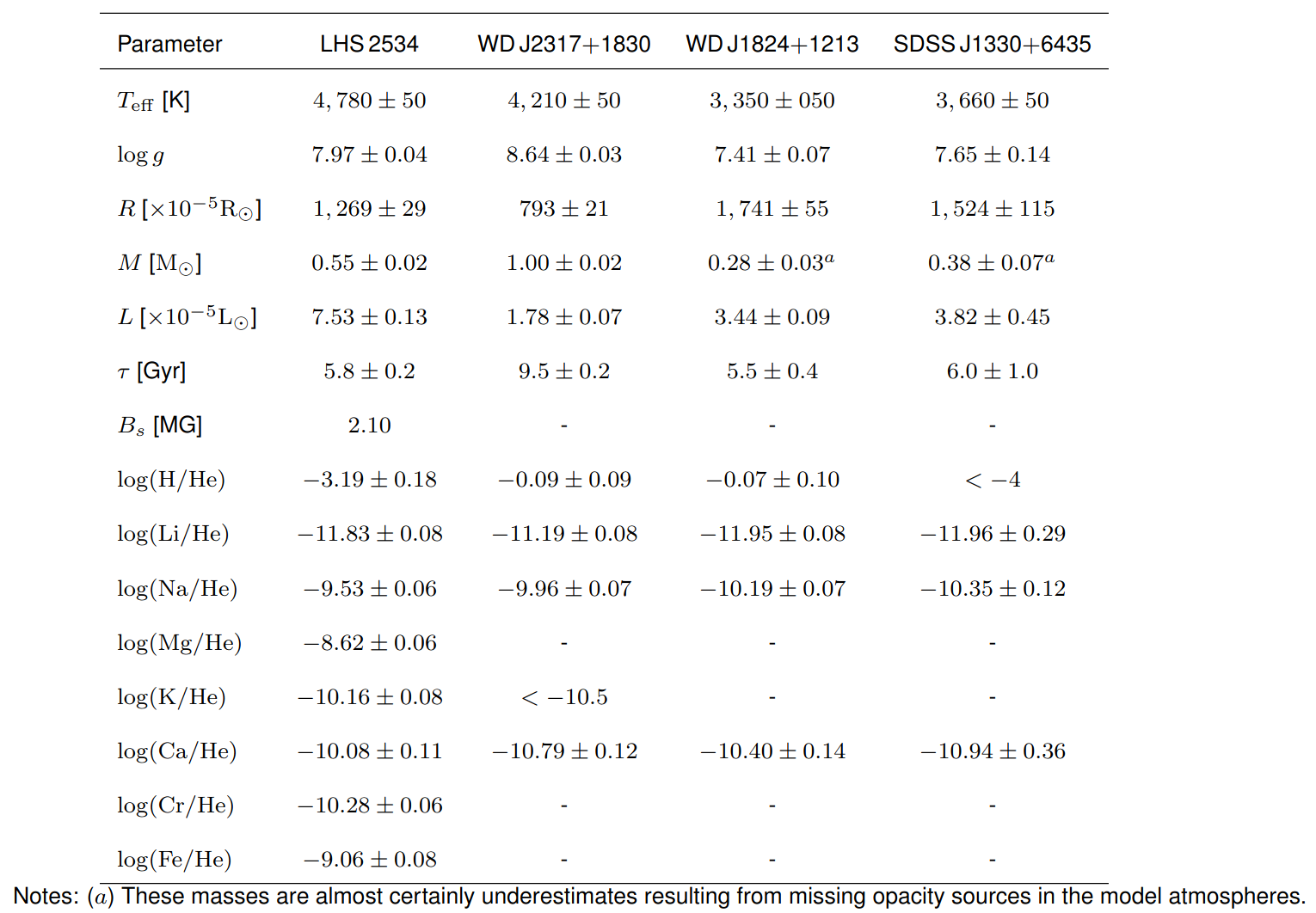}
    \caption{\label{tab:params}\textbf{Atmospheric parameters for the four white dwarfs with photospheric lithium.} The abundances are in base-10 in terms of number ratio.
    }
\end{efigure}

\begin{efigure}
  \centering
  \includegraphics[angle=0,height=0.78\textheight]{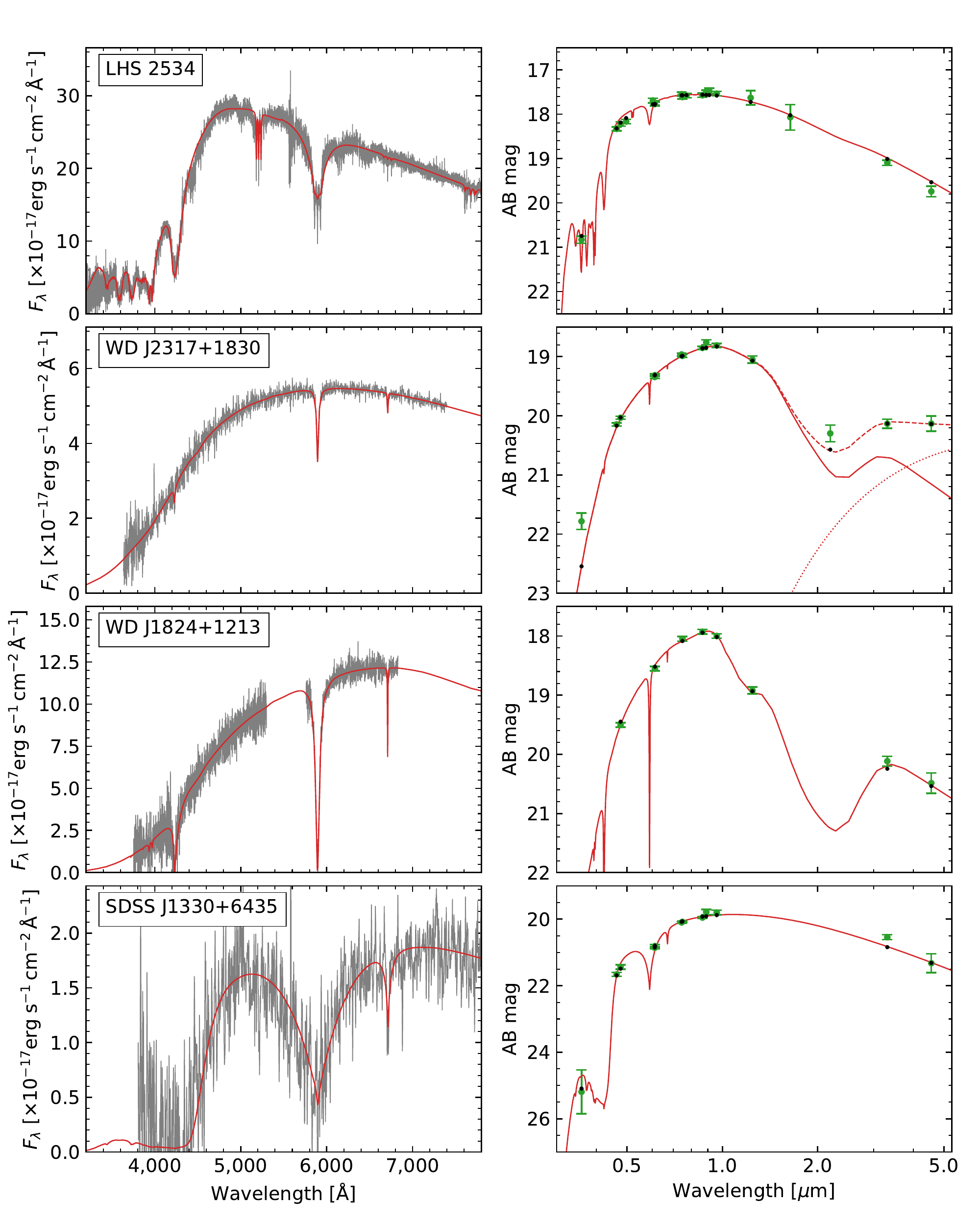}
  \caption{\label{fig:allfits}\textbf{Best fitting models compared with the spectra and photometry of
  the four lithium-bearing sample.} In the right-panel of \Lib\ the disk model and white dwarf plus
  disk model are indicated by dotted and dashed curves, respectively.
  The spectrum of \Lid\ has been smoothed with a Gaussian with a full width half maximum of 5\,\AA.
  Error bars correspond to 1$\sigma$ uncertainties.
  }
\end{efigure}

\begin{efigure}
  \centering
  \includegraphics[angle=0, scale=0.29]{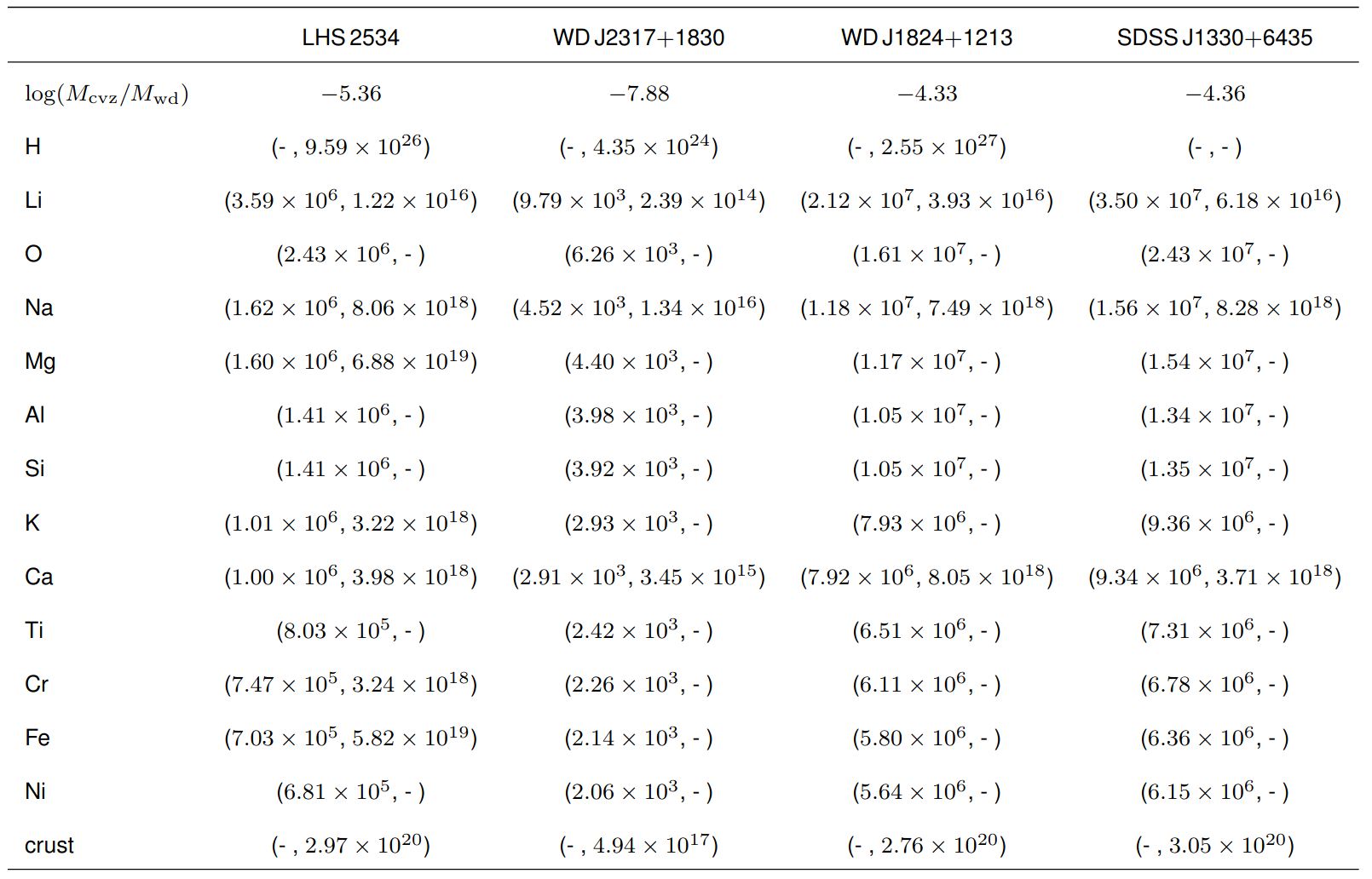}
  \caption{\label{tab:cvz}\textbf{White dwarf envelope parameters for our sample.}
  The first row indicates the fractional convection zone mass. In subsequent rows, pairs correspond to
  the sinking timescale at the base of the convection zone in years,
  and (where abundances were determined) the elemental mass in the
  convection zone in g, i.e. $(\tau_Z/\mathrm{yr}, m_Z/\mathrm{g})$. Diffusion timescales
  are given for all elements commonly considered in white dwarf planetary abundance studies.
  The final row, ``crust'', provides estimates for the total material within the white dwarf convection
  zones, assuming a continental-crust composition\cite{rudnick+gao03-1}, scaled from the Na masses.
  }
\end{efigure}

\begin{efigure}
  \centering
  \includegraphics[angle=0,width=\textwidth]{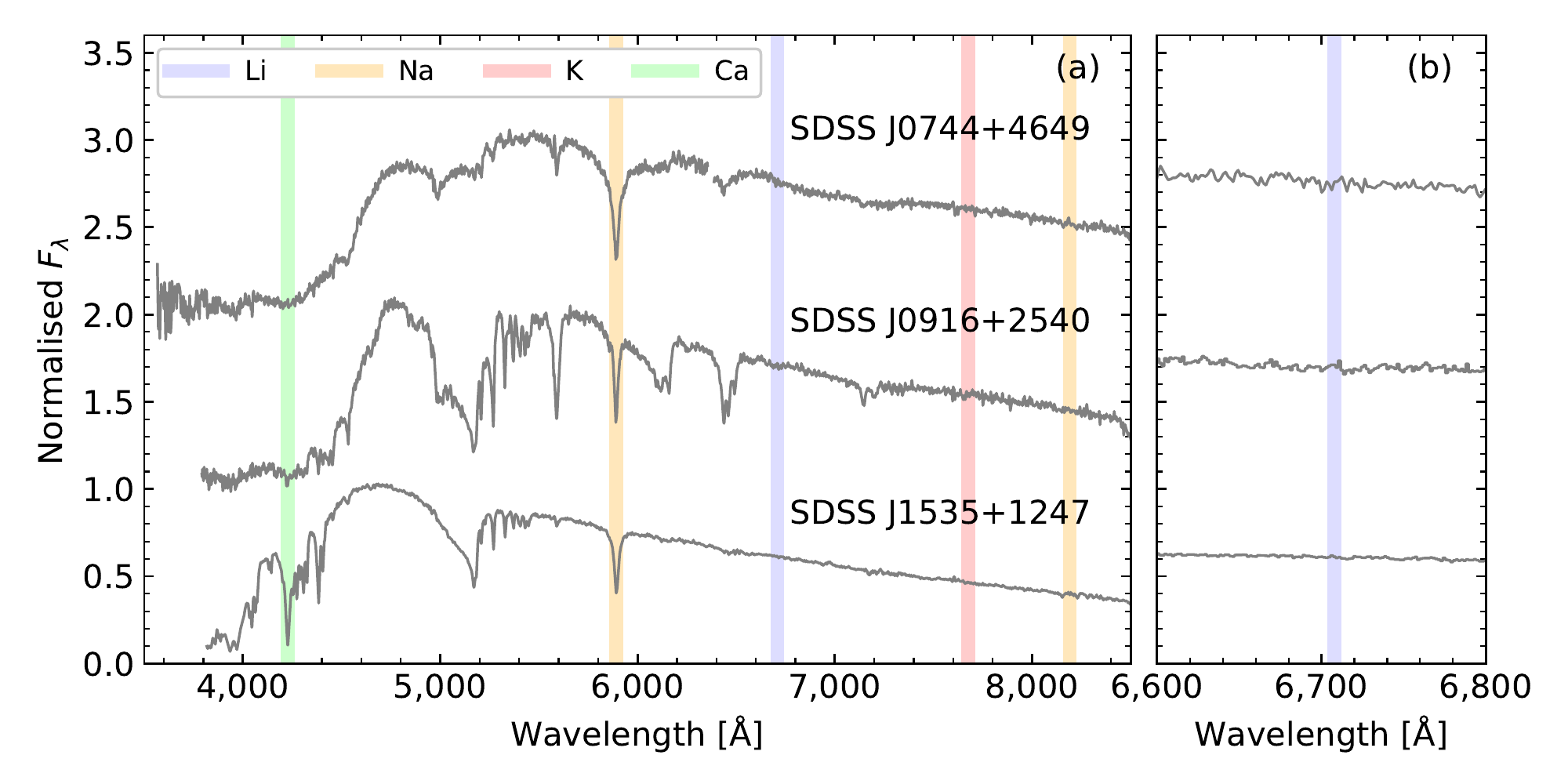}
  \caption{\label{fig:morespec}
  \textbf{SDSS spectra of three additional cool DZs with strong metal absorption features}. Lithium lines are not
  detected for any of these stars. Spectra have been smoothed by a Gaussian with a full width half maximum of 3\,\AA\ for clarity.
  }
\end{efigure}

\clearpage


\begin{sfigure}
  \includegraphics[angle=0,width=0.8\columnwidth]{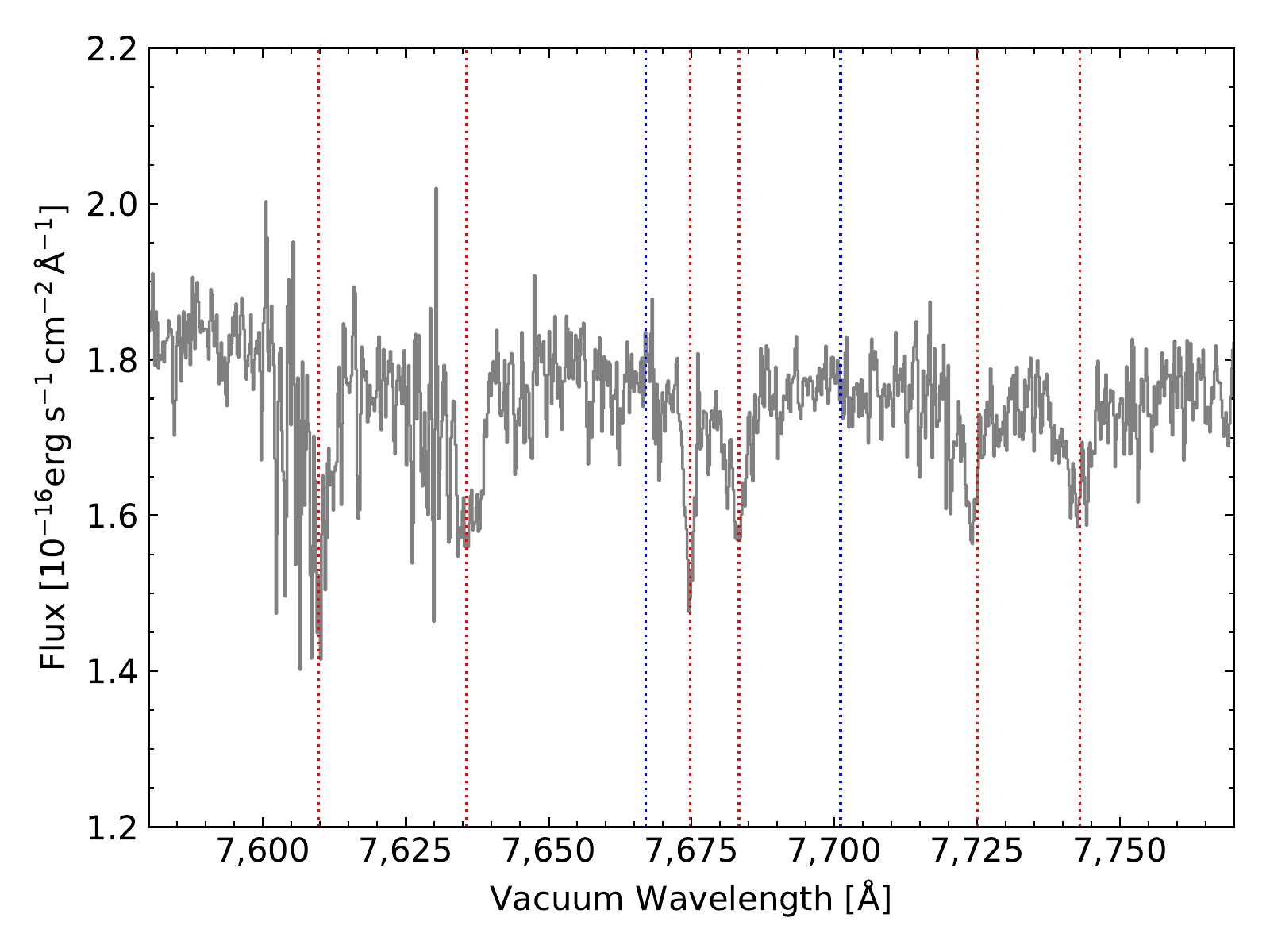}
  \caption{\label{fig:KI}\textbf{Zeeman splitting of \Ion{K}{i} at \Lia.}
  \Ion{K}{i} is observed in the regime
  intermediate to the anomalous Zeeman and Paschen-Back effects. The blue
  dotted lines indicate the expected wavelengths in the field-free case,
  whereas the red dotted lines mark the calculated wavelengths for a 2.10\,MG
  field. The spectrum wavelengths have been transformed to the rest-frame.
  }
\end{sfigure}

\begin{sfigure}
  \centering
  \includegraphics[angle=0,width=\textwidth]{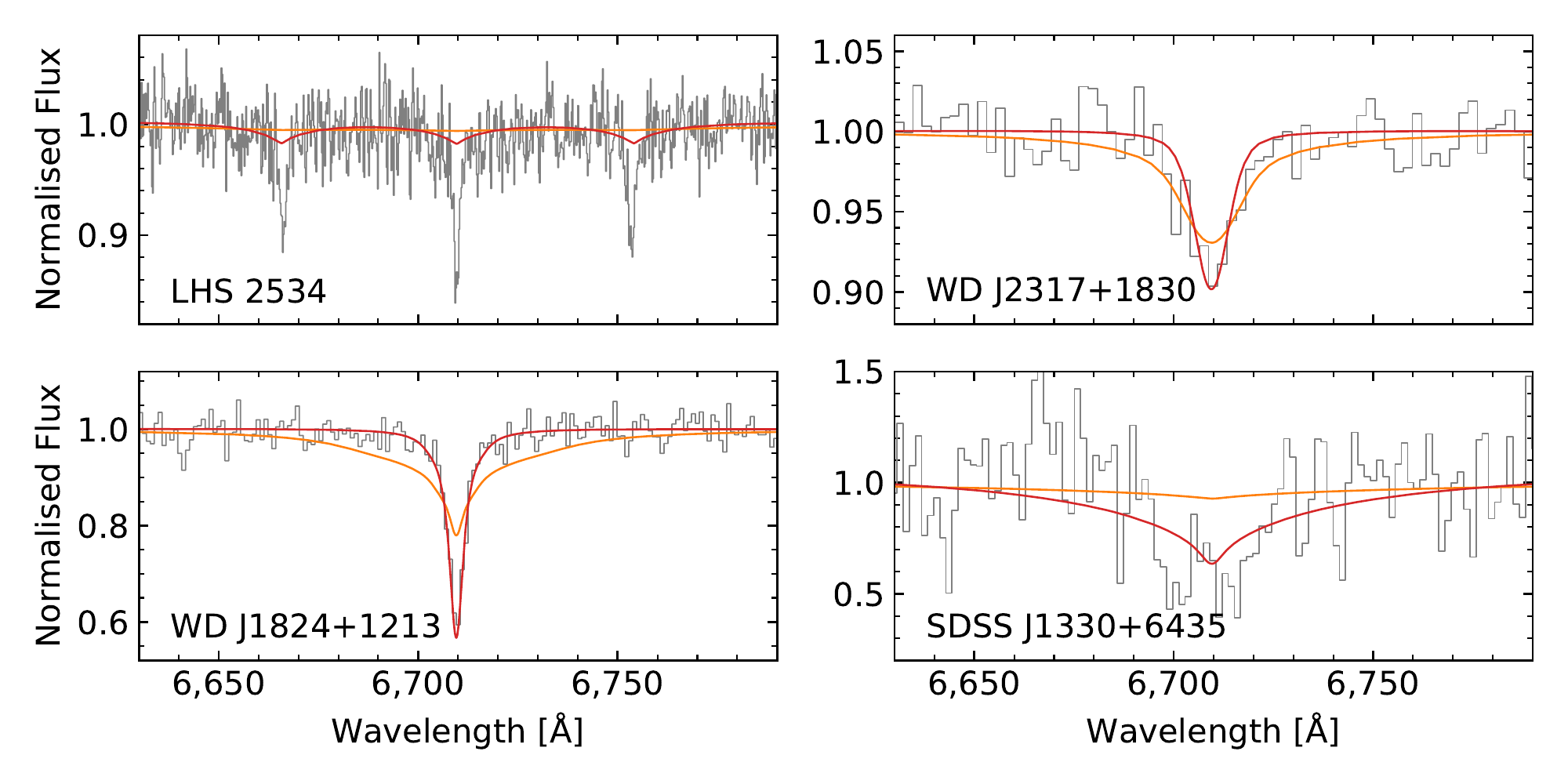}
  \caption{\label{fig:Li_lines}\textbf{Models compared with the lithium lines for each of our stars.}
  Our best fitting models,
  with the revised neutral-broadening constant are shown in red. The models with the nominal broadening
  constants are shown in orange, demonstrating the poor quality fits (note that abundances are increased
  by $+0.3$\,dex for visibility).
  }
\end{sfigure}

\begin{stable}
  \centering
  \includegraphics[angle=0,scale=0.29]{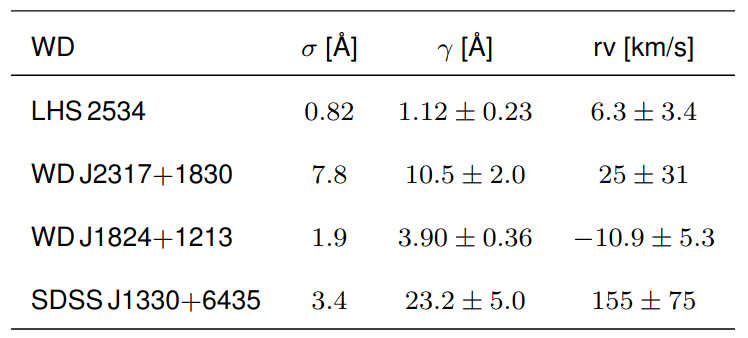}
    \caption{\label{tab:Liwidths}\textbf{Measured widths of \Ion{Li}{i} lines.}
    The spectral resolution, $\sigma$, was measured from the width of sky
    emission lines associated with the spectrum of each star. The Lorentzian
    component, $\gamma$, was determined from fitting a Voigt profile to each
    Li-doublet (the $\pi$-component in the case of \Lia).}
\end{stable}

\begin{stable}
  \centering
  \includegraphics[angle=0,scale=0.29]{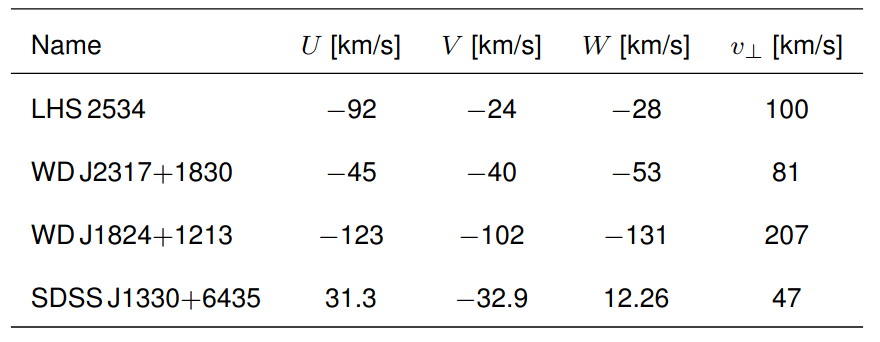}
    \caption{\label{tab:kinematics}\textbf{Tangential velocity and kinematics in the Galactic frame.}
    The coordinate $U$ is radial away from the Galactic center, $V$ is in the direction of rotation, whereas $W$ is perpendicular to the Galactic disk. We have assumed zero radial velocity.}
\end{stable}

\end{document}